\documentclass[12pt]{iopart}

\usepackage{iopams}
\usepackage{graphicx}
\usepackage{epstopdf}
\usepackage{subfig}
\usepackage{color}
\usepackage{bm}
\usepackage{citesort}

\usepackage{ulem}

\begin{document}

\title[$n = 1$ resistive wall modes in CFETR 1GW scenario]{Destabilizing effects of edge infernal components on n = 1 resistive wall modes in CFETR 1GW steady-state operating scenario}

\author{Rui HAN}
\address{Department of Plasma Physics and Fusion Engineering, School of Nuclear Science and Technology, University of Science and Technology of China, Hefei 230026, China}

\author{Ping ZHU}
\address{International Joint Research Laboratory of Magnetic Confinement Fusion and Plasma Physics, State Key Laboratory of Advanced Electromagnetic Engineering and Technology, School of Electrical and Electronic Engineering, Huazhong University of Science and Technology, Wuhan, Hubei 430074, China}
\address{Department of Engineering Physics, University of Wisconsin-Madison, Madison, Wisconsin 53706, USA}
\ead{zhup@hust.edu.cn}

\author{Linjin ZHENG}
\address{Institute of Fusion Studies, University of Texas at Austin, Austin, Texas 78712, USA}

\author{the CFETR Physics Team}

%\vspace{10pt}
%\begin{indented}
%\item[]November 2019
%\end{indented}

%\newpage
\begin{abstract}

The stability of the $n=1$ resistive wall modes (RWMs) is investigated using the AEGIS code for the newly designed China Fusion Engineering Test Reactor (CFETR) 1GW steady-state operating (SSO) scenario. Here, $n$ is the toroidal mode number. Due to the large fraction of bootstrap current contribution, the profile of safety factor q is deeply reversed in magnetic shear in the central core region and locally flattened within the edge pedestal. Consequently the pressure-driven infernal components develop in the corresponding q-flattened regions of both core and edge. However, the edge infernal components dominate the $n=1$ RWM structure and lead to lower $\beta_N$ limits than the designed target $\beta_N$ for the CFETR 1GW SSO scenario. The edge rotation is found the most critical to the stabilization due to the dominant influence of the edge infernal components, which should be maintained above $1.5\%\Omega_{A0}$ in magnitude in order for the rotation alone to fully suppress the $n=1$ RWM in the CFETR 1GW SSO scenario.

\end{abstract}

\vspace{2pc}
%

\submitto{\NF}
%\maketitle
%\newpage
\section{Introduction}
\label{sec1}

%================
% paragraph 1
%================

The external kink instability, which occurs when plasma $\beta_N$ exceeds the no-wall beta limit $\beta_N^{no-wall}$~\cite{troyon1984ppcf}, can be fully suppressed by a close perfectly conducting wall up to the higher ideal-wall beta limit $\beta_N^{ideal-wall}$. Here, $\beta_N$ is the normalized plasma $\beta$ defined as in $\beta_N=\beta_t a B_t/I_p$, where $\beta_t=2\mu_0 \langle p \rangle /B_t^2$, $\langle p \rangle$ the volume-averaged plasma pressure, $B_t$ the toroidal magnetic field, $a$ the plasma minor radius, and $I_p$ the toroidal plasma current. In reality, however, the finite wall resistivity can lead to the growth of resistive wall mode (RWM) and eventually the potential onset of plasma disruption, which is a major concern for tokamak operation and performance~\cite{freidberg1987MHD,chu2010ppcf}.

The China Fusion Engineering Test Reactor (CFETR) project aims to bridge the gap between the experimental fusion reactor ITER~\cite{aymar2002ppcf} and the future fusion power plant DEMO~\cite{zohm2013nf}. During the past few years, significant progress has been made in the CFETR conceptual physics and engineering design~\cite{wanbaonian2014ieee,songyuntao2014ieee,chan2015nf,maoshifeng2015jnm,wanyuanxi2017nf,lijiangang2019jfe,jianxiang2017nf,chenjiale2017ppcf,liuli2018nf,lizeyu2018nf,chengshikui2019ppcf,renzhenzhen2019nf,hanrui2020ppcf}. Since 2018, a new design scenario of CFETR has been adopted with major radius $R_0=7.2m$ and minor radius $a=2.2m$, primarily targeting the long-pause self-sustainable burning plasmas with fusion power production up to 1GW in the steady-state operation (SSO)~\cite{zhuangge2019nf}. 

Such SSO scenarios, in which a high bootstrap current fraction is obtained together with good energy confinement and favourable MHD stability properties, are now considered to be the most promising for future reactor scale tokamaks~\cite{joffrin2007ppcf}. However, the large fraction of bootstrap current, leads to significant reduction in the magnetic shear, and thus the associated stabilization against the MHD instabilities driven by the steep pressure gradients in the transport barrier regions~\cite{peeters2000ppcf}. In this case an pressure-driven MHD mode, the infernal mode, can appear if the zero-shear, i.e. the `flat-q' region has a safety factor value close to a low-order rational value, which might then destroy the barriers~\cite{manickam1987nf,zhenglinjin2013pop,dong2017pop}. For more common cases where the flat-q region is away from rational values, the infernal components become part of the RWM and lead to lower beta limits~\cite{zhenglinjin2017nf}. 

In earlier advanced tokamak studies, more attention is drawn to the influence of the central safety factor reversal. In D\uppercase\expandafter{\romannumeral3}-D~\cite{reimerdes2007prl} and JET experiments~\cite{reimerdes2006pop}, it is found that the no-wall beta limit for this kind of discharge is approximately $\beta_N^{no-wall} = 2.5li$, where $li$ is the plasma inductance, whereas the ratio $\beta_N^{no-wall}/li$ usually scales as $4$ for the conventional scenarios ~\cite{lao1992pof,strait1994pop,taylor1995pop}. However, for future larger tokamaks, such as CFETR, the local bootstrap current density tends to be strong enough to form another flattened safety factor profile in the edge pedestal region. In this case the edge infernal components can develop and further destabilize the RWMs. Such a destabilizing effect of edge infernal components has been confirmed in this work, where the stability of the $n=1$ RWM has been analyzed using the AEGIS code~\cite{zhenglinjin2005prl,zhenglinjin2006jcp} for the CFETR 1GW SSO scenario. Consequently, stronger edge localized toroidal rotation is found to be required for the suppression of the $n=1$ RWM, which is above 1.5\% the core Alfv\'enic speed for the 1GW design. 

The rest of paper is organized as follows. In Sec.~\ref{sec2}, the equilibria of the CFETR 1GW SSO scenario are briefly reviewed. In Sec.~\ref{sec3} the numerical scheme of AEGIS code is introduced. In  Sec.~\ref{sec4} the stability analysis of the $n=1$ RWMs for the CFETR 1GW SSO scenario is reported, as well as the influence of edge infernal components on the $\beta_N$ limits. Sec.~\ref{sec5} evaluates the rotational stabilization effects, including those of rotation profiles. Conclusions and discussions are given in Sec.~\ref{sec6}.\\

%=====================================

\section{Equilibrium profiles and parameters of CFETR 1GW SSO scenario} 
\label{sec2}

\begin{figure}[htbp] 
\centering

\includegraphics[width=0.55\textwidth]{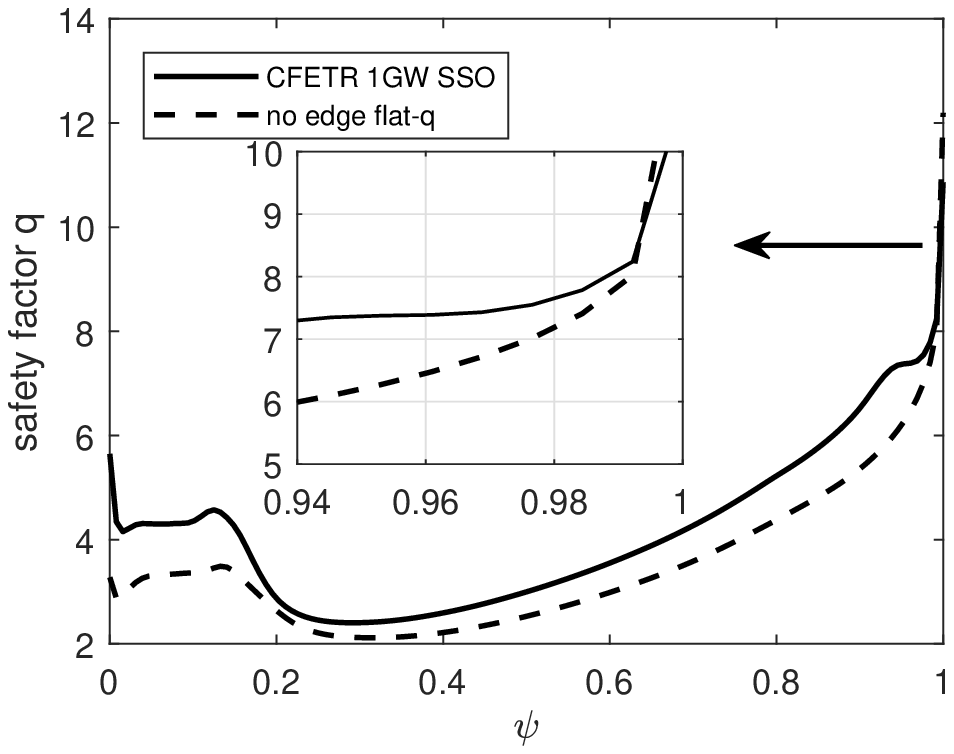}
\put(-245,160){\textbf{(a)}}\\
\includegraphics[width=0.55\textwidth]{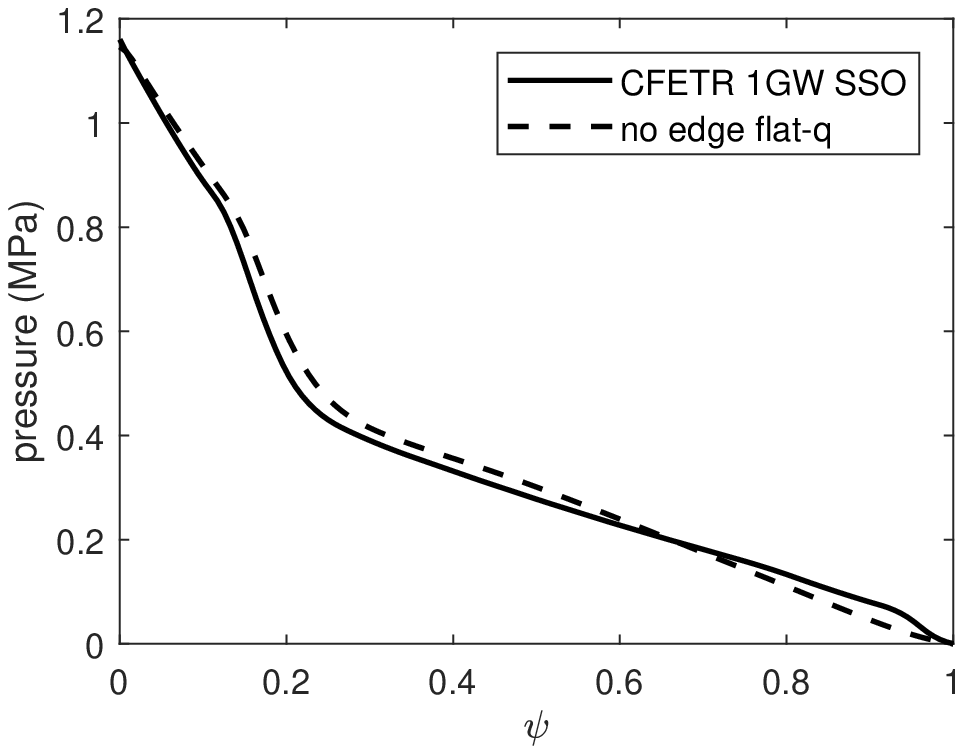}
\put(-245,160){\textbf{(b)}}\\
\includegraphics[width=0.55\textwidth]{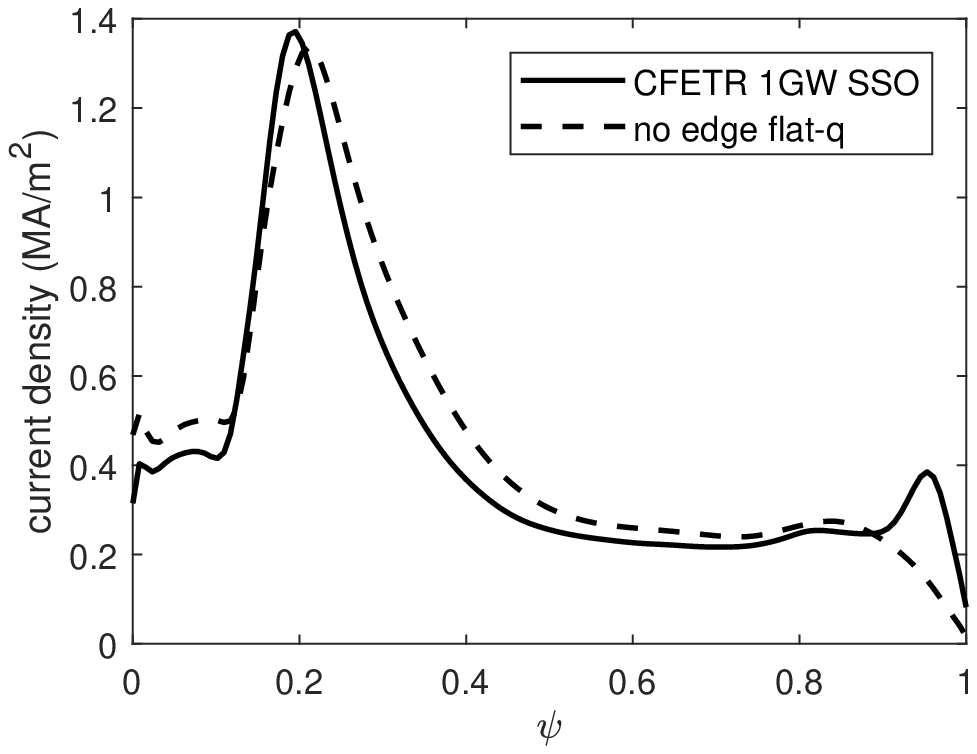}
\put(-245,160){\textbf{(c)}}

\caption{(a) Safety factor, (b) pressure, and (c) current density profiles as functions of the normalized magnetic flux $\psi$ in the CFETR 1GW SSO scenario equilibrium (solid lines) and the `no edge flat-q' equilibrium (dashed lines).}
\label{fig_eq} 
\end{figure}

The CFETR 1GW SSO scenario is designed based on the integrated modelling in the OMFIT framework, where the EFIT code is employed to construct the MHD equilibrium used in this study~\cite{jianxiang2017nf,chenjiale2017ppcf}. The main parameters are as follows: major radius $R_0=7.2m$, minor radius $a=2.2m$, toroidal magnetic field at magnetic axis $B_0=6.53T$, total plasma current $I_p = 11 MA$, plasma inductance $li=0.75$, the normalized beta $\beta_N=2.86$, and the Alfv\'enic speed at magnetic axis $v_A=1.004\times10^7 m/s$. The current density profile is locally peaked at $\psi=0.18$ in the core and $\psi=0.96$ at edge (figure~\ref{fig_eq}). Correspondingly, the safety factor has a reversal in magnetic shear in the central region and a flat region in the edge pedestal ranging from $\psi=0.94$ to $\psi=0.99$. For comparison, a reference equilibrium is prepared by reducing the bootstrap current and pressure gradient in the edge pedestal region, so that the flat q profile at edge region is removed as shown in figure~\ref{fig_eq} with dashed lines. The total current and core plasma pressure are kept same as the SSO case.

%=====================================

\section{Ideal MHD model in AEGIS code}
\label{sec3}

\begin{figure}[htbp]
\centering
\includegraphics[width=0.8\textwidth]{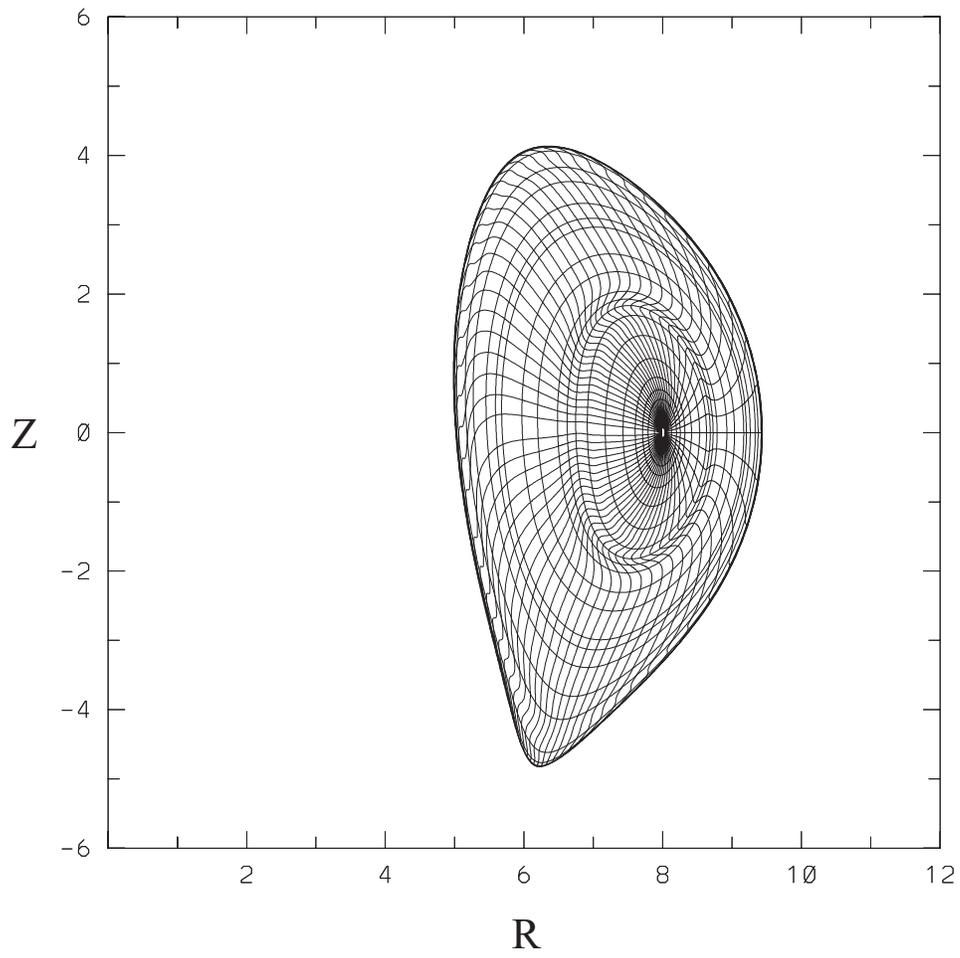}
\caption{AEGIS computational domain and mesh for the CFETR SSO scenario equilibrium.}
\label{fig_grid} 
\end{figure}

The numerical studies in this work are carried out using the AEGIS code~\cite{zhenglinjin2005prl,zhenglinjin2006jcp}. The following linearized ideal MHD equation for toroidal plasma is solved:
\begin{equation}
-\rho_m \hat{\omega}^2\bm{\xi}=\delta\bm{J}\times\bm{B} + \bm{J}\times\delta\bm{B}-\nabla\delta P
\label{eq_aegis}
\end{equation}
where $\rho_m$ is the mass density, $\bm{\xi}$ the perpendicular fluid displacement, $\bm{J}$ the equilibrium current density, $\bm{B}$ the equilibrium magnetic field, $P$ the equilibrium pressure, $\mu_0\delta\bm{J}=\nabla\times\delta\bm{B}$, $\delta\bm{B}=\nabla\times(\bm{\xi}\times\bm{B})$, and $\delta P=-\bm{\xi}\cdot\nabla P$. The rotation effects are included mainly through the Doppler shift, $\hat{\omega}=\omega+n\Omega$ with $\Omega$ being the toroidal rotation frequency. For the subsonic rotation regime in most tokamak plasmas, the effects from centrifugal and Coriolis forces are negligible~\cite{waelbroeck1991pfb,zhenglinjin1999pop}. In addition, since the RWM frequency in this study is much smaller than the ion acoustic wave frequency, the plasma compressibility results only in the so-called apparent mass effect through the coupling between parallel and perpendicular inertia motions~\cite{greene1962pf}. Therefore, the contribution of plasma compressibility can be included by regarding $\rho_m$ as the apparent mass in the above momentum equation~(\ref{eq_aegis}). To avoid the singularity in edge safety factor near the separatrix, the diverted equilibria of CFETR used in this study are all truncated at the poloidal flux surfaces where the safety factor value $q_a =9.1$ (figure~\ref{fig_grid}). The conformal shaped wall is used in our AEGIS calculations.

%=====================================

\section{Influence of edge infernal components on $\beta_N$ limits}
\label{sec4}

\begin{figure}[htbp]
\centering
\includegraphics[width=0.8\textwidth]{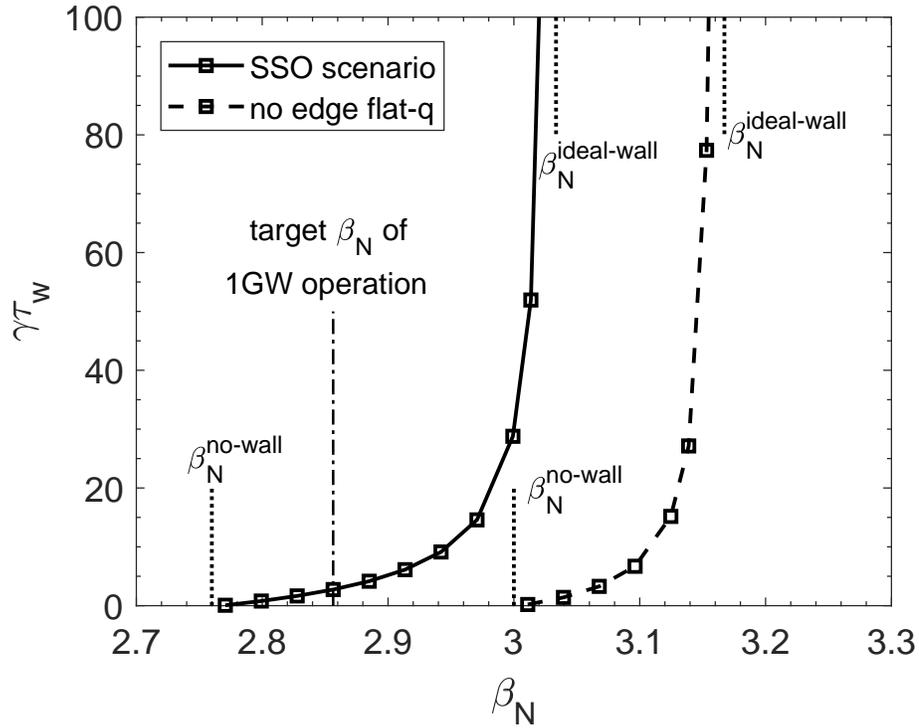}
\caption{The $n=1$ RWM growth rates as functions of normalized beta $\beta_N$ for the CFETR 1GW SSO scenario equilibrium (solid line) and the `no edge flat-q' equilibrium (dashed lines) in presence of a resistive wall at $r_w=1.2a$, with $\tau_w=7.17ms$. The no-wall and ideal-wall beta limits are marked with shorter vertical dotted lines. The target $\beta_N$ of CFETR 1GW SSO scenario is marked with the longer vertical dash-dotted line.}
\label{fig_beta} 
\end{figure}

In this section we analyze the linear stability of $n=1$ RWM for the CFETR 1GW SSO scenario, as well as the influence of flattened q-profile in the edge pedestal region. We compare the RWM stability of the SSO scenario and the reference equilibrium in absence of the flattened q-profile near the edge as described in section~\ref{sec2} (figure~\ref{fig_beta}),  where $\tau_w$ is the resistive diffusion time of the wall, typically a few milliseconds. For the designed wall location at $r_w=1.2a$, we find that the reference equilibrium without the edge flattened q-profile has higher beta limits ($[\beta_N^{no-wall}, \beta_N^{ideal-wall}]=[3, 3.17]$) than the SSO case ($[\beta_N^{no-wall}, \beta_N^{ideal-wall}]=[2.76, 3.03]$). This indicates that the edge flat q-profile is unfavourable for RWM stability. The target $\beta_N$ of CFETR 1GW SSO scenario is $2.86$, which falls within the unstable RWM regime when none of the passive or active stabilizing mechanisms is taken into account.  

\begin{figure}[htbp]
\centering
\begin{minipage}{0.45\textwidth}
\includegraphics[width=1.0\textwidth]{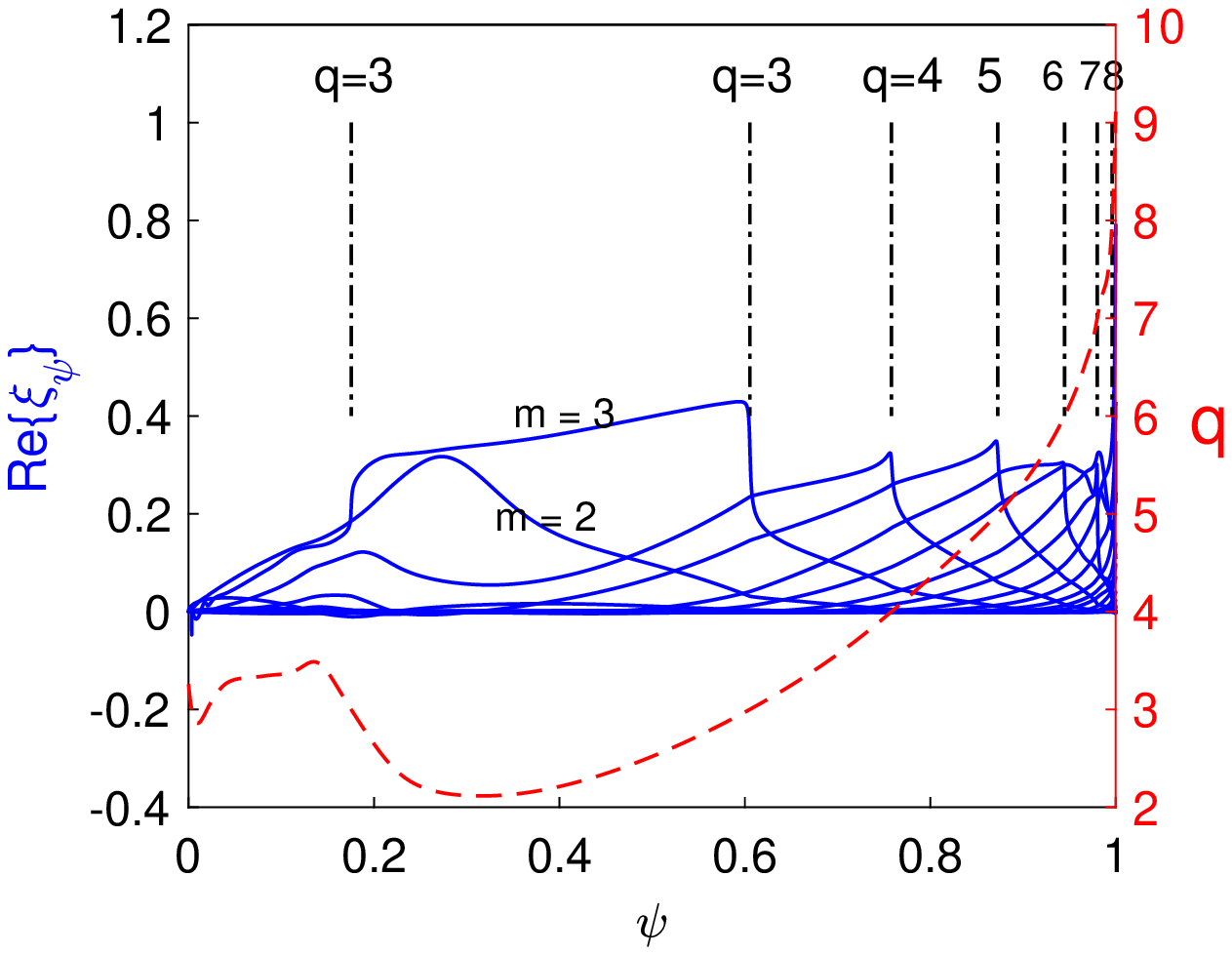}
\put(-170,130){\textbf{(a)}}
\end{minipage}
\begin{minipage}{0.45\textwidth}
\includegraphics[width=1.0\textwidth]{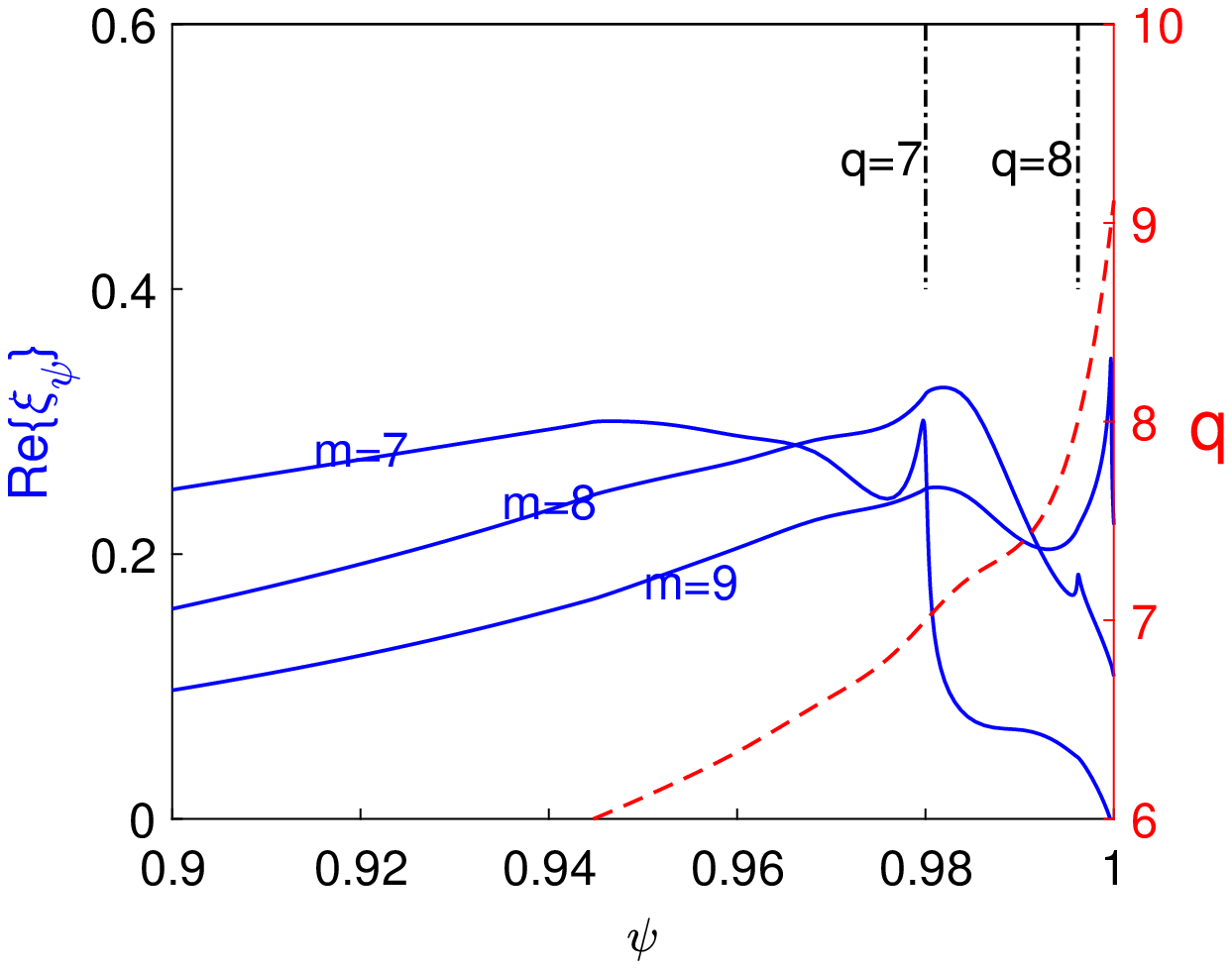}
\put(-170,130){\textbf{(b)}}
\end{minipage}
\caption{(a) Real components of the radial displacement of $n=1$ RWM as functions of the normalized magnetic flux $\psi$ with wall position $r_w =1.2a$ for the `no edge q-flat' equilibrium with $\beta_N=3.07$. (b) The radial profiles of $m=7-9$ harmonics in the edge region. The safety factor profile is plotted as the red dashed lines.}
\label{fig_case2_eigen} 
\end{figure}

\begin{figure}[htbp]
	\centering
	\begin{minipage}{0.45\textwidth}
		\includegraphics[width=1.0\textwidth]{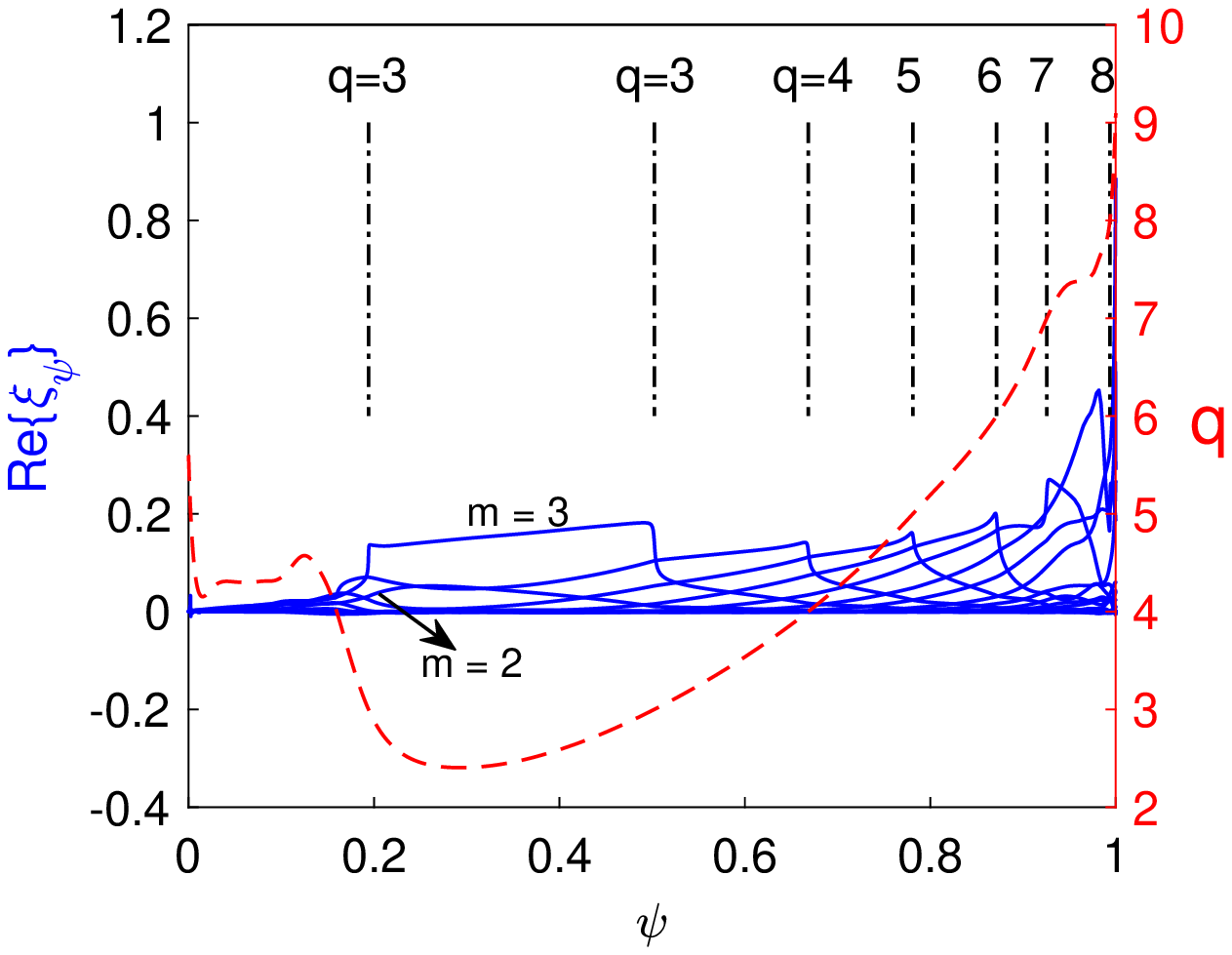}
		\put(-170,130){\textbf{(a)}}
	\end{minipage}
	\begin{minipage}{0.45\textwidth}
		\includegraphics[width=1.0\textwidth]{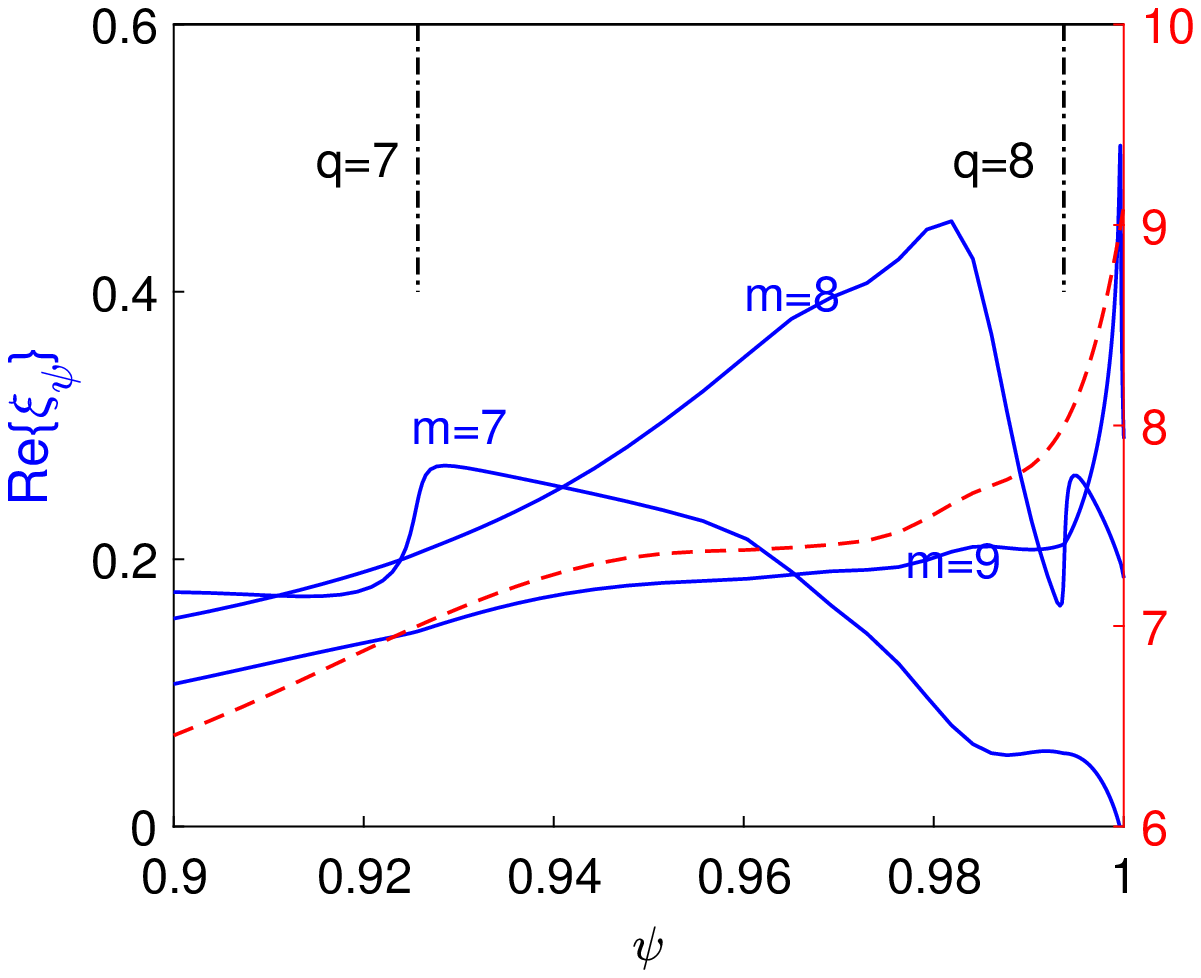}
		\put(-170,130){\textbf{(b)}}
	\end{minipage}
	\caption{(a) Real components of the radial displacement of $n=1$ RWM as functions of the normalized magnetic flux $\psi$ with wall position $r_w =1.2a$ for the CFETR 1GW SSO equilibrium. (b) The radial profiles of $m=7-9$ harmonics in the edge region. The safety factor profile is plotted as the red dashed lines.}
	\label{fig_sso_eigen} 
\end{figure}

The presence of edge flat q region can significantly change the RWM mode structure. In the reference equilibrium case where the edge flattened q region is absent (figure~\ref{fig_case2_eigen}), the $m=2$ harmonic peaks around the central flat q region. This is unusual since the minimum $q$ value is above 2 thus the $m=2$ harmonic is expected to be non-resonant. Meanwhile, the $m=3$ harmonic broadly domes over the region between the two $q=3$ rational surfaces with a greater amplitude. For the SSO case (figure~\ref{fig_sso_eigen}), however, the edge infernal component with poloidal number $m=8$ becomes more dominant, even though the pressure gradient as well as the current density in the edge region is weaker than that in the central region, where the $m=2$ and $3$ harmonics are much reduced. In particular, unlike all other harmonics which peak around rational surfaces, the $m=8$ harmonic peaks in the edge flattened q region where $q_{flat}=7.38$ and $\psi=0.98$, which is also crucial to the rotation effects on the RWMs (see next section). Thus the edge flat q profile is more influential than the central flat q profile on the RWM mode structure and stability. This may be because that the RWM is an external MHD mode, so that the plasma current in the edge region has larger impact than that in the central region. 

\begin{figure}[htbp]
	\centering
	\begin{minipage}{0.45\textwidth}
		\includegraphics[width=1.0\textwidth]{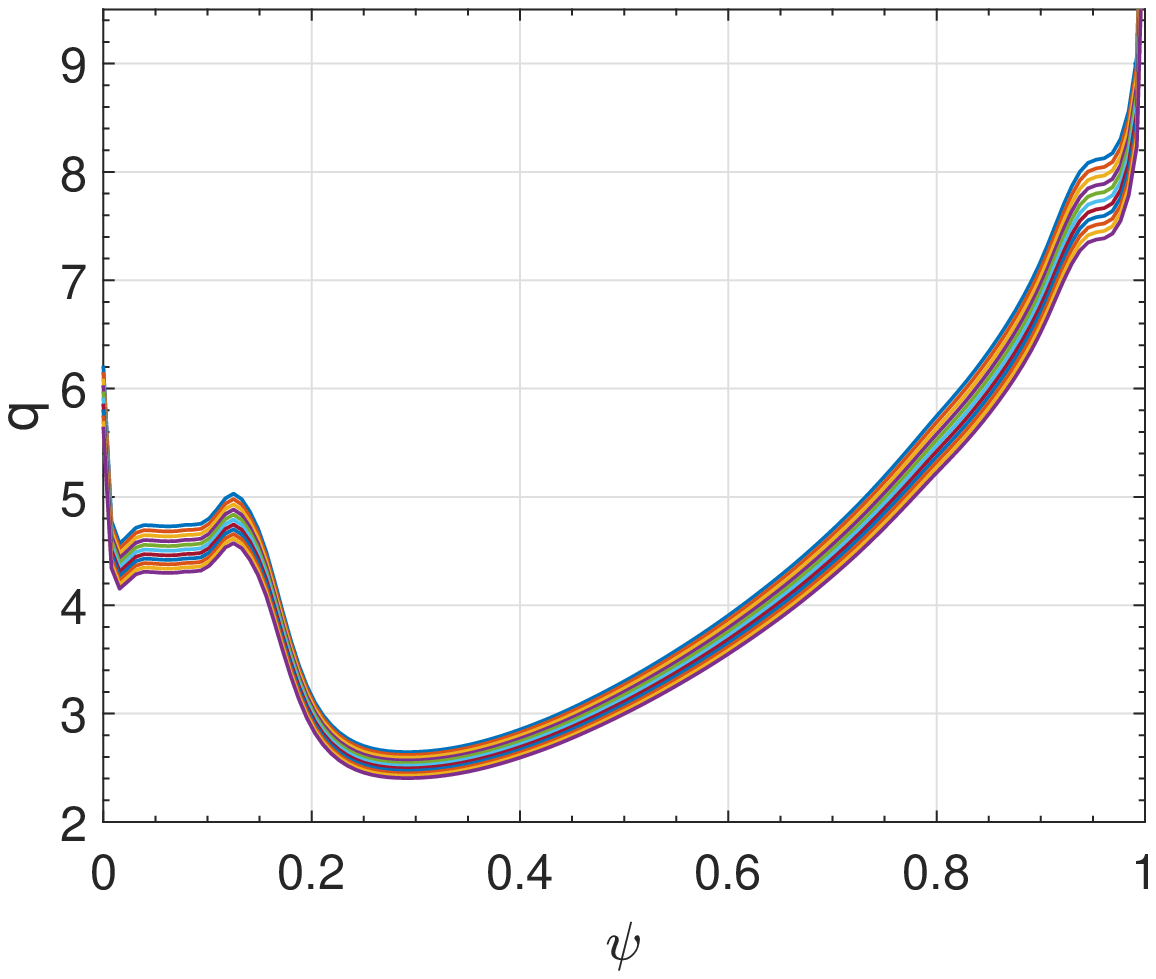}
		\put(-170,130){\textbf{(a)}}
	\end{minipage}
	\begin{minipage}{0.45\textwidth}
		\includegraphics[width=1.0\textwidth]{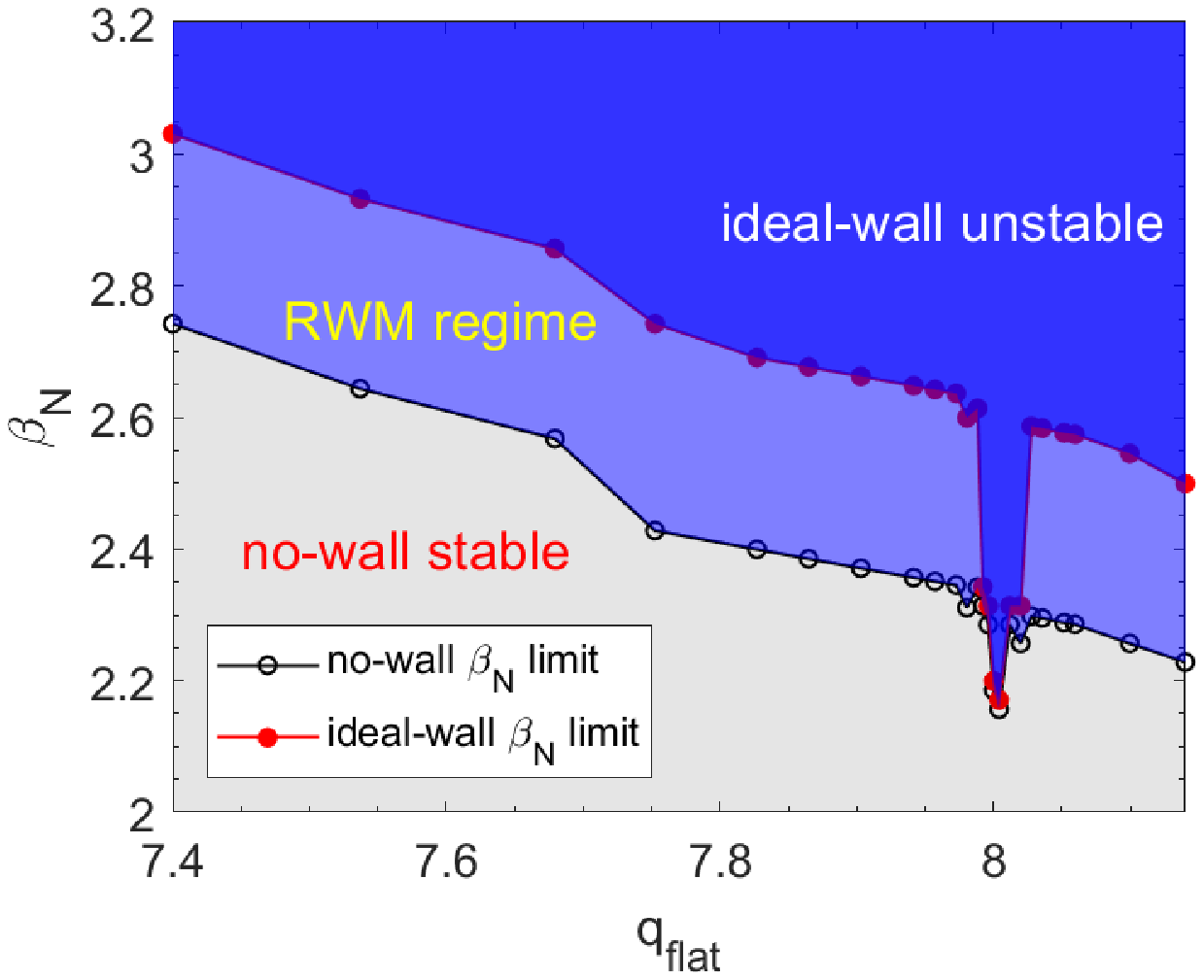}
		\put(-170,130){\textbf{(b)}}
	\end{minipage}
	\caption{(a) Safety factor profiles for a series of equilibria with increasing safety factor values. (b) The no-wall and ideal-wall $\beta_N$ limits as functions of $q_{flat}$.}
	\label{fig_qmod} 
\end{figure}

Next, the influence of safety factor value at the edge flat q region on the stability is studied. We slightly vary the total toroidal current of the SSO scenario equilibrium while keeping the shape of current density profile same. A series of equilibria with $q_{flat}$ ranging from $7.4-8.2$ are generated as shown in figure~\ref{fig_qmod}(a). The stability of $n=1$ modes for these equilibria are evaluated. Figure~\ref{fig_qmod}(b) shows the no-wall and ideal-wall $\beta_N$ limits as functions of the $q_{flat}$. One can see that both $\beta_N$ limits dramatically drop when $q_{flat} \simeq 8$, whereas apart from that they gradually decrease with $q_{flat}$. Moreover, both $\beta_N$ limits become nearly identical at $q_{flat}=8$, indicating the complete loss of wall stabilization. Note that in this case, the eigenfunction becomes delta-function-like and peaks at the resonance point for a single Fourier component, which can be identified as the infernal mode. Thus, the presence of the edge flat q region may subject the SSO scenario to the $n=1$ external instability even at a considerably low $\beta_N$.

\section{Rotational stabilization effects on RWM with edge infernal components}
\label{sec5}

In this section, we evaluate the stabilization effects on RWM due to toroidal plasma rotation with three types of radial profiles: the uniform rotation profile, the radially descent profile, and the Gaussian rotation profile. 

\subsection{Uniform profile}

\begin{figure}[htbp]
\centering
\begin{minipage}{0.45\textwidth}
	\includegraphics[width=1.0\textwidth]{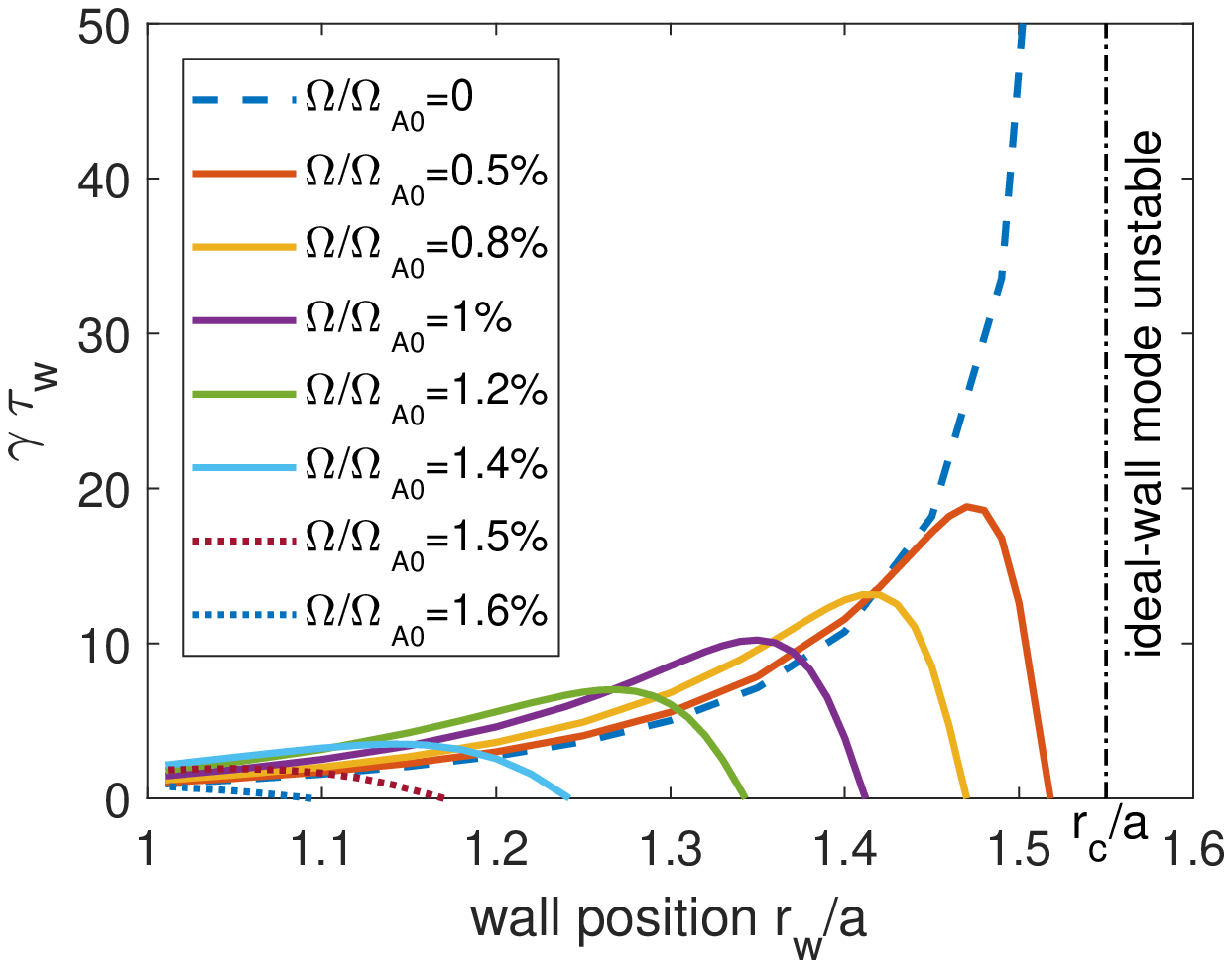}
	\put(-100,130){\textbf{(a)}}
\end{minipage}
\begin{minipage}{0.45\textwidth}
	\includegraphics[width=1.0\textwidth]{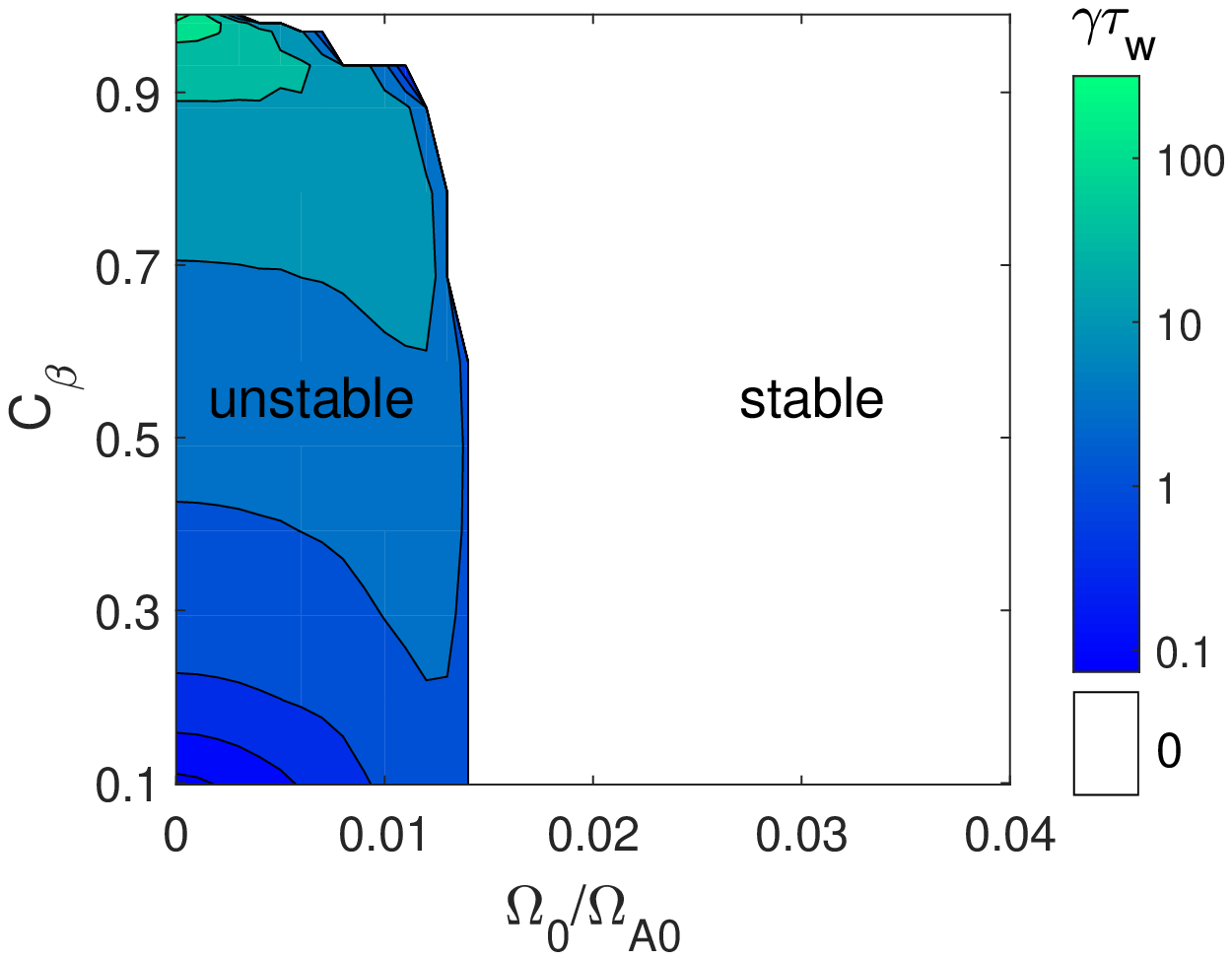}
	\put(-100,130){\textbf{(b)}}
\end{minipage}
\caption{(a) The $n=1$ RWM growth rates $\gamma\tau_w$ as functions of wall position $r_w/a$ in presence of uniform rotations with various frequencies $\Omega/\Omega_{A0}$ for the CFETR 1GW SSO scenario with fixed $\beta_N=2.86$. The critical wall position $r_c=1.55a$ is plotted as the vertical dashed line. (b) RWM growth rates $\gamma\tau_w$ in 2D parameter space ($C_{\beta}$, $\Omega/\Omega_{A0}$) with fixed wall position $r_w=1.2a$.}
\label{fig_uniform_rot} 
\end{figure}

We first compare the RWM growth rate variation over the first-wall location for a range of rotation frequencies accessible in the SSO scenario with the target $\beta_N = 2.86$ (figure~\ref{fig_uniform_rot}(a)). In absence of rotation, the RWM growth rates tend to infinity near the critical wall position $r_c=1.56a$. In presence of rotation, the RWM growth rates initially increase with the wall radius, but then drop quickly to zero beyond a certain wall radius $r_b$. Since $r_b < r_c$, as a result, a `stable window' opens up between $r_b$ and the critical wall position $r_c$. The stronger the rotation, the wider the `stable window' $r_c - r_b$, which can be used as a measurement of the rotational stabilization effects. For the designed wall position $r_w=1.2a$, the RWM can be fully suppressed by a rotation frequency of $\Omega/\Omega_{A0}=1.5\%$, where $\Omega_{A0}$ is the Alfv\'enic frequency defined as $\Omega_{A0} = v_A / R_0$. Figure~\ref{fig_uniform_rot}(b) shows the RWM stability diagram in the 2D parameter space ($C_{\beta}$, $\Omega/\Omega_{A0}$) with the wall position $r_w=1.2a$. Here, $C_{\beta}=(\beta_N-\beta_N^{no-wall})/(\beta_N^{ideal-wall}-\beta_N^{no-wall})$ is a scaling factor ranging from $0-1$ between the two $\beta_N$ limits. It is found that a uniform rotation with $\Omega/\Omega_{A0}=1.5\%$ provides sufficient stabilization on RWM for all achievable $\beta_N$.

\subsection{Radially descent profile}

\begin{figure}[htbp]
	\centering
	\includegraphics[width=0.8\textwidth]{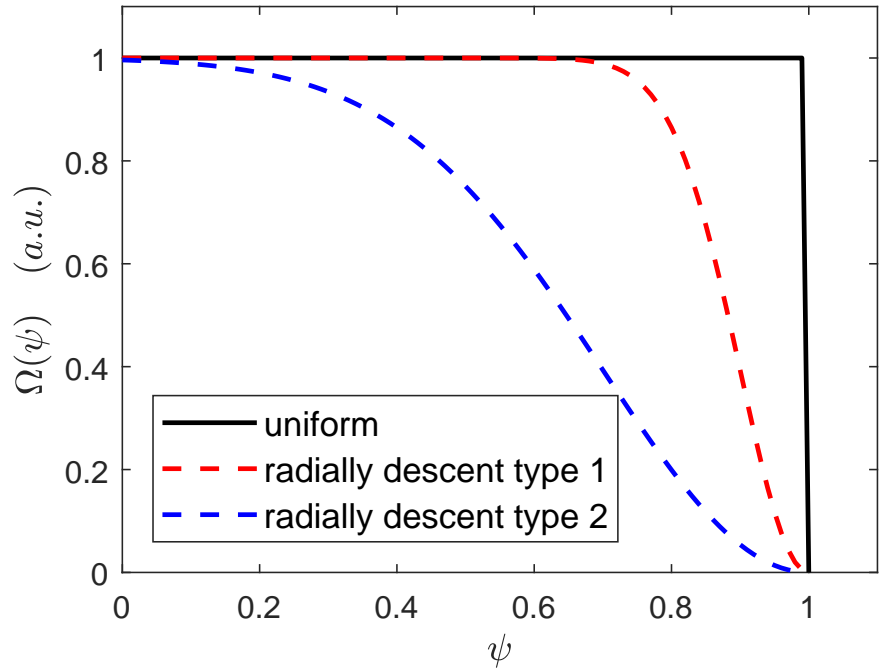}
	\caption{Toroidal rotation frequencies as functions of the normalized magnetic flux $\psi$.}
	\label{fig_profile_descent} 
\end{figure}

\begin{figure}[htbp] 
\centering
\begin{minipage}{0.45\textwidth}
\includegraphics[width=1.0\textwidth]{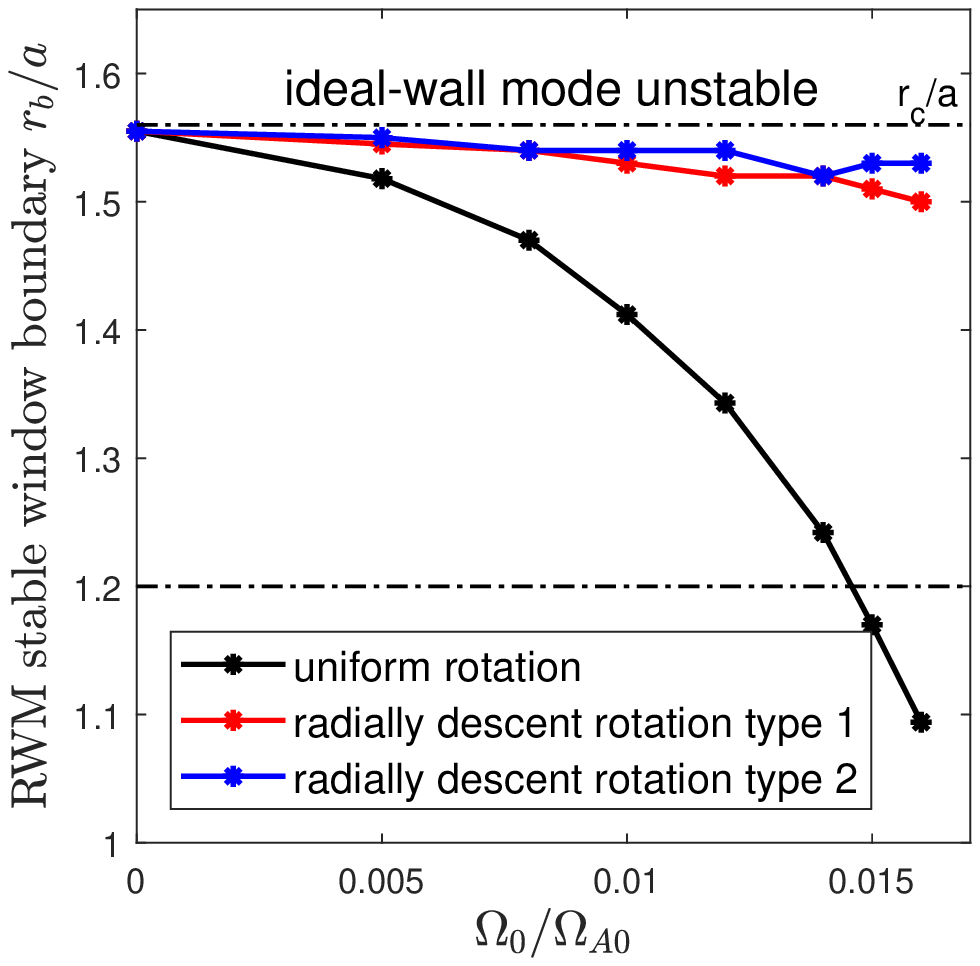}
\put(-170,175){\textbf{(a)}}
\end{minipage}
\begin{minipage}{0.45\textwidth}
\includegraphics[width=1.0\textwidth]{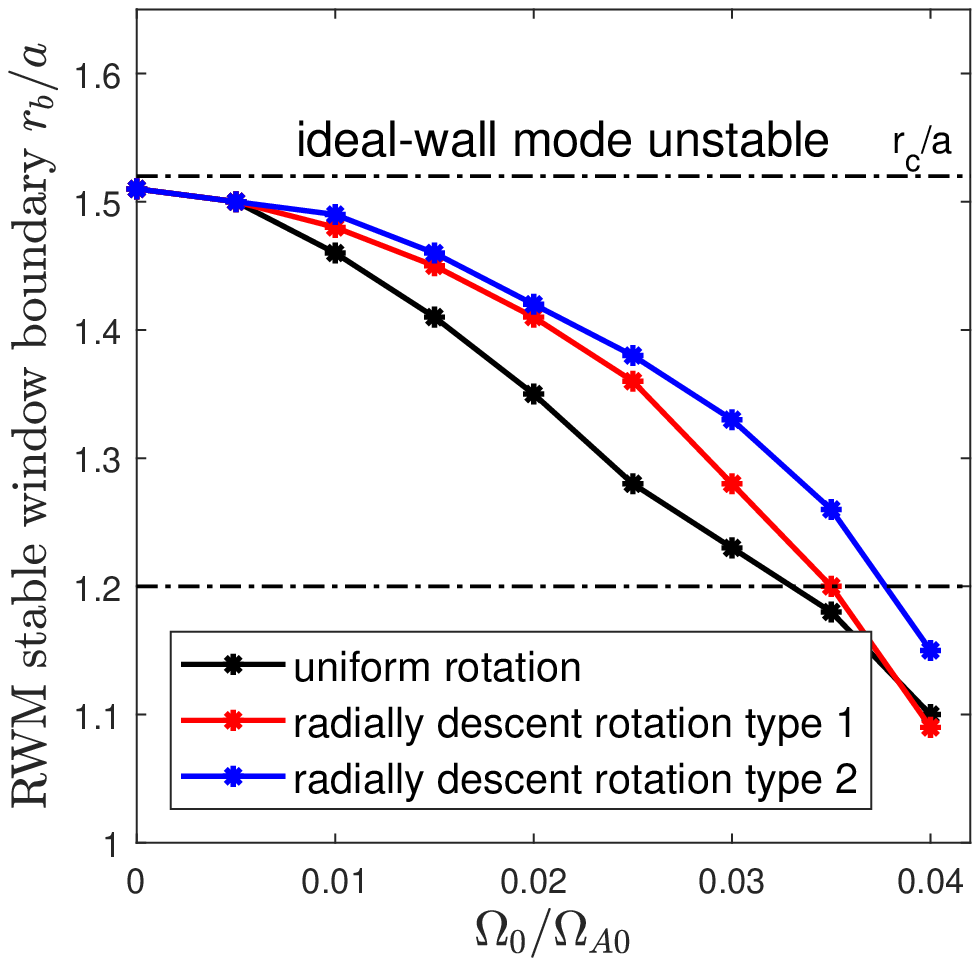}
\put(-170,175){\textbf{(b)}}
\end{minipage}
\caption{The lower boundaries of RWM 'stable window' $r_b/a$ as functions of the rotation frequency at magnetic axis $\Omega_0/\Omega_{A0}$ for three different rotation profiles in (a) the CFETR 1GW SSO scenario equilibrium with $\beta_N=2.86$ and (b) the `no edge flat-q' equilibrium with $\beta_N=3.07$.}
\label{fig_stabilization_descent} 
\end{figure}

In this subsection, we consider a toroidal rotation with nearly uniform radial profile in the region of $\psi=0$ to $0.6$ and rapidly descent profile in the region of $\psi=0.6$ to $1$, which is similar to the design for ITER~\cite{chrystal2017pop}; and a rotation with gradually descent profile in the entire radial direction, which is similar to the design for CFETR~\cite{chenjiale2017ppcf}. The rotation profiles are plotted as `type 1' and `type 2' in figure~\ref{fig_profile_descent}. The reference equilibrium without the edge flat q region is also considered for comparison (figure \ref{fig_stabilization_descent}). With radially descent profiles, rotation can barely open up the `stable window' for the SSO case, whereas the `stable window' can readily open up and become wider with larger rotation frequency in the reference case. This suggests that the toroidal rotation with radially descent profile may not be effective or sufficient for stabilizing the RWM in the SSO scenario due to the strongly destabilizing presence of the edge infernal mode component localized in the edge flat q region.

\subsection{Gaussian profile}

\begin{figure}[htbp] 
\centering
\begin{minipage}{0.45\textwidth}
\includegraphics[width=1.0\textwidth]{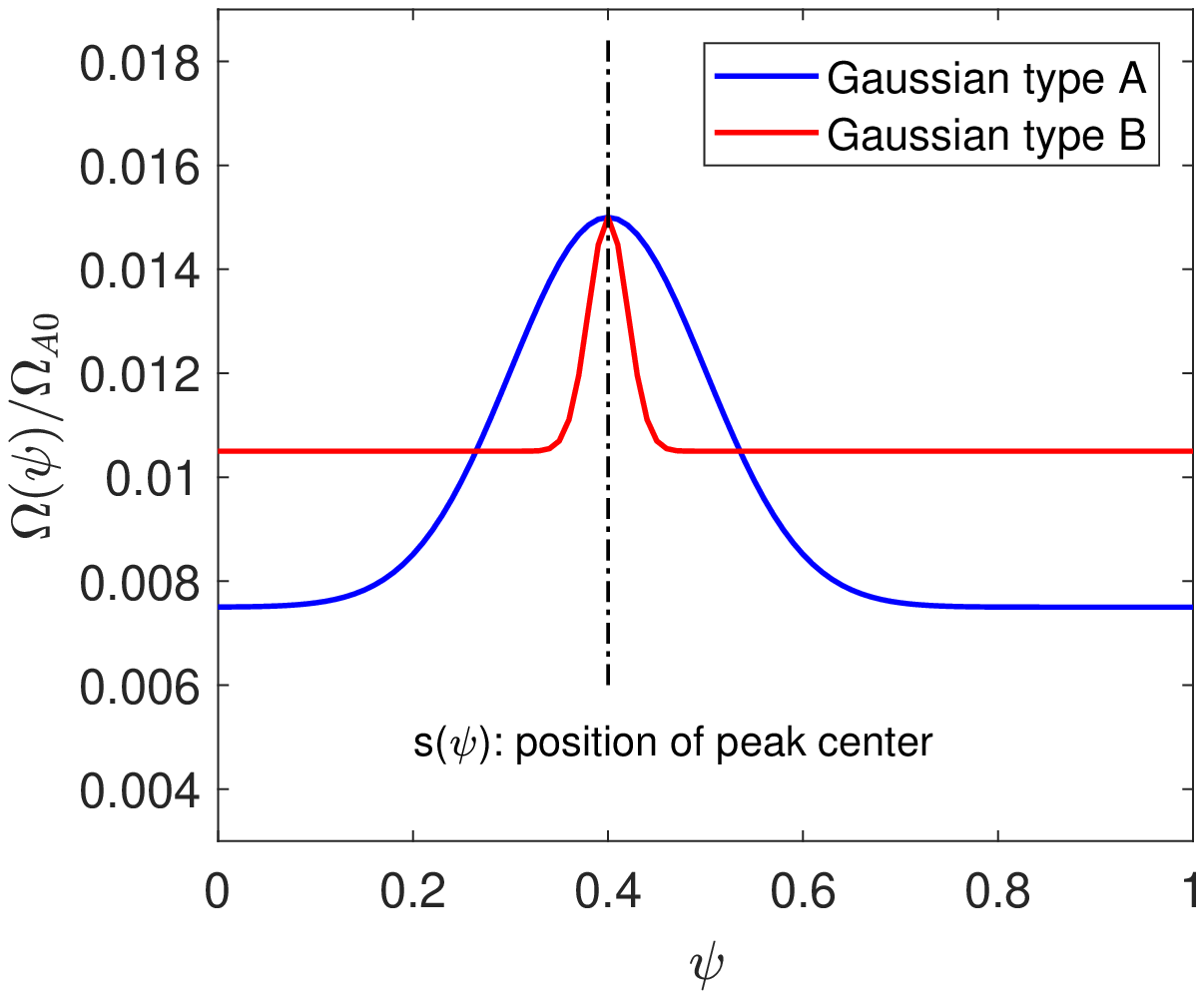}
\put(-150,135){\textbf{(a)}}
\end{minipage}
\begin{minipage}{0.45\textwidth}
\includegraphics[width=1.0\textwidth]{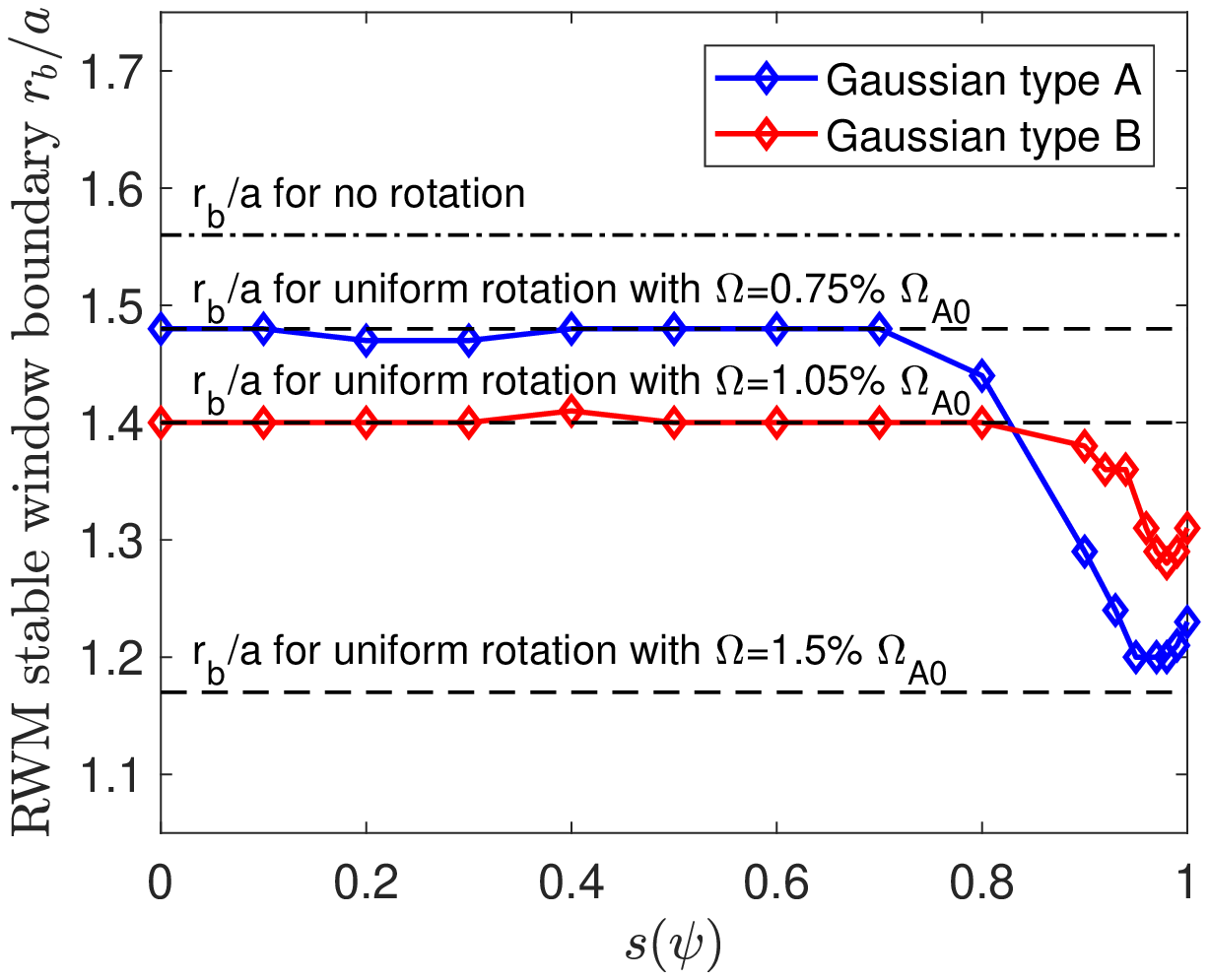}
\put(-150,135){\textbf{(b)}}
\end{minipage}
\caption{(a) Gaussian rotation profiles of: type A rotation with a broad positive peak and type B rotation with a narrow positive peak. $\Omega_0=1.5\%\Omega_{A0}$. The setup of shape parameters $[A_0,A_1,\sigma]$ for type A rotation is $[0.5,0.5,0.1]$ and type B rotation $[0.7,0.3,0.02]$. (b) The lower boundaries of RWM `stable window' $r_b/a$ as functions of the rotation peak position $s$ for the CFETR 1GW SSO scenario. The lower stability boundaries $r_b/a$ for uniform rotations with $\Omega/\Omega_{A0}=0$, $0.75\%$, $1.05\%$ and $1.5\%$ are plotted as the constant horizontal dashed lines respectively for comparisons.}
\label{fig_rot_gaussian} 
\end{figure}

We further consider the Gaussian shaped rotation profiles as shown in figure~\ref{fig_rot_gaussian}(a) and parameterized as follows,

\begin{equation*}
	\Omega(\psi)=\Omega_0(A_0+A_1e^{-\frac{(\psi-s)^2}{2\sigma^2}})
\end{equation*}
where the peak location $s$ ranges from 0 to 1. We investigated the effects of these rotation profiles on RWM `stable window' for the CFETR 1GW SSO scenario. Figure~\ref{fig_rot_gaussian}(b) shows the lower boundary of the RWM `stable window' $r_b/a$ as functions of rotation peak position $s$. When the rotation profiles are peaked in the core region (i.e. $ s < 0.6$), their stabilizing effects are not different from the uniform rotation with the same corresponding off-peak rotation frequency, since they share the same RWM `stable window', respectively. However, when the peak of the profile moves to plasma edge, the boundaries of the `stable window' change significantly. The maximum stabilizing effects are obtained when $s=0.98$, where the $m=8$ edge infernal component peaks (figure~\ref{fig_sso_eigen}). When the edge peaked rotation profile is also sufficiently broad to cover the region of edge flat q-profile or infernal mode component, a peak rotation frequency above $1.5\%\Omega_{A0}$ can fully stabilize the RWM in the CFETR 1GW SSO scenario.

%=====================================
\section{Summary and Discussions}
\label{sec6}

In summary, we have studied the $n=1$ RWM stability in the CFETR 1GW SSO scenario using the AEGIS code. Due to the large bootstrap current contribution, such an SSO scenario has a deep safety factor reversal in the core region and a flat q profile in the edge region, corresponding to the internal and external transport barriers, respectively. As a consequence, the infernal mode components develop in both core and edge regions. The edge infernal component dominates the destabilizing effects for RWMs, which lower the $\beta_N$ limits and render the $n=1$ RWM unstable in the CFETR 1GW SSO scenario. 

The $n=1$ RWMs can be fully suppressed by uniform rotation with frequency above $1.5\%\Omega_{A0}$. Due to the influence of the edge infernal components, it is the edge rotation that mainly contributes to the stabilization effect. Thus an edge localized toroidal rotation with a sufficiently broad profile and its peak rotation frequency maintained at above $1.5\%\Omega_{A0}$ can fully stabilize the $n=1$ RWMs in the CFETR 1GW SSO scenario. However, a robust control of RWM in CFETR SSO may require a full account of all other possible stabilizations from kinetic effects, as well as the active stabilizations from the feedback control coil system, which should be evaluated in other and future work.

\ack

This work is supported by the National Magnetic Confinement Fusion Science Program of China Grant No. 2019YFE03050004, the National Natural Science Foundation of China Grant Nos. 11775221 and 51821005, the National Key Research and Development Program of China No. 2017YFE0300500, 2017YFE0300501, the Fundamental Research Funds for the Central Universities at Huazhong University of Science and Technology Grant No. 2019kfyXJJS193, and the U.S. DOE Grant Nos. DE-FG02-86ER53218 and DE-SC0018001. This research used the computing resources from the Supercomputing Center of University of Science and Technology of China.

%=====================================

\section{References}
\bibliography{cfetr_ss}

\newpage
%=====================================
% Fig.1
%=====================================

\newpage
%=====================================
% Fig.2
%=====================================

\newpage
%=====================================
% Fig.3
%=====================================

\newpage
%=====================================
% Fig.4
%=====================================

\newpage
%=====================================
% Fig.5
%=====================================
\newpage

%=====================================
% Fig.6
%=====================================

\newpage
%=====================================
% Fig.7
%=====================================

\newpage
%=====================================
% Fig.8
%=====================================

\newpage
%=====================================
% Fig.9
%=====================================

\newpage
%=====================================
% Fig.10
%=====================================

\end{document}